\documentclass[pra,aps,twocolumn,showpacs]{revtex4}

\usepackage{graphicx}% Include figure files

\usepackage{bm}% bold math
\usepackage{amsmath}
\usepackage{amsfonts}
\usepackage{amssymb}
\usepackage{color}

\newcommand{\ke}[1]{|#1\rangle}

\begin{document}
\pacs{37.10.Mn,37.30.+i}
\title{Cavity sideband cooling of trapped molecules}

\author{Markus Kowalewski,$^1$ Giovanna Morigi,$^{2,3}$ Pepijn W.H. Pinkse,$^4$ Regina de Vivie-Riedle$^1$}

\affiliation{$^1$ Department of Chemistry, Ludwig-Maximilian-Universit\"at, D-81377 Munich, Germany,\\
$^2$ Departament de F\'{\i}sica, Universitat Aut\`onoma de Barcelona,
E-08193 Bellaterra, Spain,\\
$^3$ Theoretische Physik, Universit\"at des Saarlandes, D-66041 Saarbr\"ucken, Germany\\
$^4$ MESA+ Institute for Nanotechnology, University of Twente, P.O. Box 217, 7500AE Enschede, The Netherlands}

\begin{abstract}
The efficiency of cavity sideband cooling of trapped molecules is theoretically investigated for the case where the IR
transition between two rovibrational states is used as a cycling transition. The molecules are assumed to be trapped
either by a radio-frequency or optical trapping potential, depending on whether they are charged or neutral, and
confined inside a high-finesse optical resonator which enhances radiative emission into the cavity mode. Using realistic
experimental parameters and COS as a representative molecular example, we show that in this setup cooling to the trap
ground state is feasible.
\end{abstract}

\date{\today}

\maketitle

\section{Introduction}

\begin{figure}[b]
\begin{center}
\includegraphics[width=0.45\textwidth]{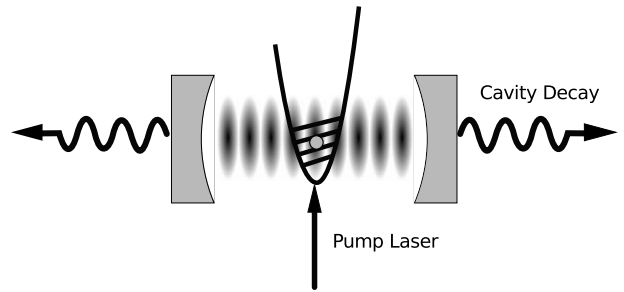}
\caption{Schematic representation of the setup. A molecule is confined by an external harmonic potential inside a
high-finesse resonator. The transition between two molecular vibrational levels is driven transversally by a laser and
scatters photons into a quasi-resonant cavity mode. The scattering processes are tailored in order to enhance the
transitions which cool the external motion.}
\label{Fig:Setup}
\end{center}
\end{figure}

With the field of cold and ultracold molecules coming of age in the past few years, efficient cooling methods are
becoming more and more valuable. Optical methods have the advantage that they do not rely on collisional processes which
are hard to predict and which often open severe collisional loss channels. A number of ingenious optical cooling schemes
for molecules have been devised: It has been realized, e.g., that translational cooling of molecules might be possible
with
light sources with a specifically designed set of lines \cite{Bahns1996,Nguyen11}. Alternatively, the almost-closed
level scheme
a molecule like CaF has been exploited \cite{Shumann2010} to laser cool molecules as it is usually done with
\cite{DiRosa04}. Even a 3D scheme has been proposed resembling the popular magneto-optical trap for atoms
\cite{Stuhl2008}.
Zeppenfeld {\it et al.}  have proposed to use electro-optical forces to carry away up to 1K per spontaneously decaying infrared photon \cite{Zeppenfeld2009}. Optical cavities have been predicted \cite{Horak1997,Vuletic2000} and used for atoms \cite{Maunz2004} as a way to achieve translational cooling without relying on spontaneous emission, in principle
allowing cooling of molecules. In Refs. \cite{Morigi2007,Kowalewski2007} we have shown that cavity-enhanced Raman
scattering
can enable one to cool both the internal and the external degrees of freedom of molecules, using simulations based on
ab-initio calculations for OH and NO radicals. In this proposal the cavity enhances the anti-Stokes, cooling transitions
over other heating transitions.  In a recent paper Vuleti{\'c} and co workers demonstrated sideband cooling of the
motion of a single trapped ion to the ground state of the trap potential via cavity enhanced scattering
\cite{Leibrandt2009}. In the present article we make the next step and demonstrate theoretically cavity-enhanced
sideband cooling of molecules and molecular ions in a strongly confining external potential, in the setup sketched in
Fig. \ref{Fig:Setup} . We show that cooling to the ground state of the external potential can be achieved for
experimentally feasible setups. The new scheme thus offers the possibility to reach lower final temperatures compared to
the previously reported cooling  schemes \cite{Vuletic2000,Morigi2007}, where the  limit is given by the
cavity linewidth.

This article is organized as follows. In Sec. \ref{Sec:2} the theoretical model is introduced and the derivation of the
rate equations for cooling the external motion of the molecules is sketched. Moreover, the method used for obtaining the
numerical results is described. In Sec. \ref{Sec:3}  the results are reported and discussed, while the conclusions are
drawn in Sec. \ref{Sec:4}. 

\section{The model}
\label{Sec:2}

\begin{figure}
\includegraphics{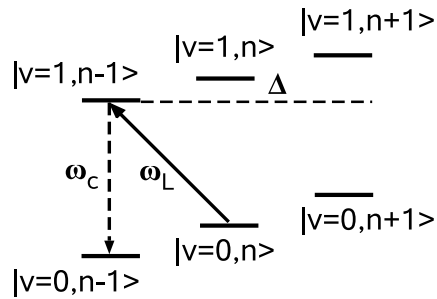}
\caption{Schematic representation of the scattering process leading to cooling. The molecular vibrational levels are
labeled by $v=0,1$, the phononic, trap excitations by $n$. The pump laser at frequency $\omega_L$  drives resonantly the
red sideband transition $|v=0,n\rangle \to |v=1,n-1\rangle$. The laser photon is scattered into a cavity mode at
frequency $\omega_c=\omega_0$, which couples resonantly with the molecular transition. Since the photon is then lost via
cavity decay, the molecule is cooled by one phonon.}
\label{Fig:Setup:2}
\end{figure}

We theoretically explore the efficiency of performing sideband cooling of a molecule of mass $M$, which is confined by a
harmonic potential and couples to the mode of a high-finesse resonator. The level scheme is sketched in
Fig. \ref{Fig:Setup:2}: a two-level transition of the molecule is driven transversally by a laser with coupling strength
$\Omega$ (Rabi
frequency), and emits photons at frequency $\omega_c$. The bad-cavity limit is assumed, which implies that the molecule
scatters prevalently photons into the cavity mode, which are then lost via cavity decay.

Let us now provide a specific example. The two-level transition, which is driven by electromagnetic radiation, is
composed of two vibrational states $v=0$, $v=1$ of the electronic ground state of a molecule, which we denote by
$|g\rangle$ and $|e\rangle$, respectively. Let $\omega_0$ be the transition frequency. This level scheme is closed
inside
the vibrational subspace. Spontaneous emission into different rotational substates is suppressed by the cavity,
while any residual leakage can be handled by microwave re-pump schemes. These properties allow one to use this
transition for cavity cooling. 

We further note that  this sort of vibrational transitions is usually in the infrared regime (100-4000\,cm$^{-1}$) and
has typically a homogeneous linewidth below 1\,kHz, which allows one to spectrally resolve the motional sidebands. In
order to obtain convenient cooling rates, and hence efficient cooling, spontaneous emission is here enhanced by
coupling with the mode of an infrared high-finesse resonator. The spatial confinement can be obtained by means of dipole
or ion traps, depending on whether the molecule is neutral or charged \cite{Ostendorf06,Drewsen00,Eijkelenborg99}. 

In order to perform sideband cooling, the initial spatial localization of the molecule must be
smaller than the wavelength of the driving laser. This requirement can be rewritten as
$\eta\sqrt{2\langle n\rangle+1} \ll 1$, with $\langle n\rangle$ the mean occupation of phononic states of the
harmonic oscillator and $\eta$ the Lamb-Dicke parameter, $\eta=\sqrt{\hbar k^2/2M\nu}$, with $k$ the wave vector of the
transition and $\nu$ the trap frequency in angular units\cite{Stenholm86,Eschner03}. One necessary condition is that the
Lamb-Dicke parameter is smaller than unity $\eta\ll 1$. Moreover, the molecule should be pre-cooled, e.g., by means of
an optical cooling schemes~\cite{Horak1997,Vuletic2000} or other non-optical methods \cite{Zeppenfeld2009}. Nevertheless
we note that sideband cooling can be efficient even for $\eta\sim 1$~\cite{Morigi97}. Moreover, schemes for ground-state
cooling for Lamb-Dicke parameters larger than unity have been theoretically proposed in~\cite{Morigi97}, and can be
extended to situations in which dissipation occurs via cavity decay.

\subsection{Master equation for cavity sideband cooling}

We now introduce the model, from which we extract the cooling rates and the final temperature expected by cavity
sideband cooling the molecules. Below we assume the Lamb-Dicke regime,
and hence expand in first order in the small parameter $\eta$~\cite{Stenholm86,Eschner03}. The model has been extensively reported, for instance in Refs.~\cite{Zippilli05,Zippilli07}.

We study the coupled dynamics of the center-of-mass motion of the molecule, the two-level transition, and the cavity mode in the reference frame rotating at the frequency $\omega_L$ of the laser, driving the two-level molecular transition. Denoting by $\rho$ the density matrix for molecular internal and external degrees of freedom (in this reduced Hilbert space) and for the cavity mode, its dynamics is governed by the master equation
\begin{equation}
\label{Eq:master}
\partial_t \rho = \frac{1}{{\rm i} \hbar}\left[ H, \rho \right] + {\cal L}_{\kappa} \rho + {\cal L}_\gamma \rho\,,
\end{equation}
where the Hamiltonian $H$ accounts for the coherent evolution, and the superoperators ${\cal L}_{\kappa}$ and ${\cal
L}_\gamma$ describe cavity decay and spontaneous emission, respectively. We now give the specific form of each term. The
full Hamiltonian can be written as the sum of the terms
\begin{equation}
\label{H}
H=H_{\rm M}+H_{\rm T}+H_{\rm C}+V_I,
\end{equation}
where
\begin{equation}
H_{M}  = - \hbar \Delta |e\rangle\langle e|
\end{equation}
is the Hamiltonian for the internal degrees of freedom, with $\Delta = \omega_L - \omega_0$. The Hamiltonian $H_{\rm T}$ describes the dynamics of the harmonic oscillator for the center-of-mass motion, which is here assumed to be one dimensional for convenience,
\begin{equation}
\label{H:trap}
H_{T} = \frac{P^2}{2M}+\frac{1}{2}M\nu^2X^2=\hbar \nu \left( b^\dagger b + \frac{1}{2} \right)\,,
\end{equation}
with $X$ and $P$ position and canonical conjugate momentum of the molecule, and $b=({\rm i}P+M\nu X)/\sqrt{2\hbar m\nu}$
annihilates a quantum of motional excitations  $\hbar\nu$ such that $[b,b^{\dagger}]=1$. We denote by $\{\ke{n}\}$ the
number states with $n=0,1,2,\ldots$, which are eigenstates of the Hamiltonian in Eq. (\ref{H:trap}).  The Hamiltonian
for the cavity mode is
\begin{equation}
H_{C}  = - \hbar \delta_c a^\dagger a\,,
\end{equation}
where operators $a$ and $a^{\dagger}$ annihilate and create, respectively, a cavity photon at frequency $\omega_c$
and $\delta_c = \omega_L - \omega_c$ is the detuning between laser and cavity mode.

In the bad-cavity regime we may reduce the Hilbert space of the cavity mode to the states $\ke{0_c}$ and $\ke{1_c}$,
namely, zero and one photon in the cavity mode, respectively. The coupling of the molecule with the laser and with the
cavity mode are given by operator $V_I$ which is here reported in the Rotating Wave Approximation (RWA), as the pump is
assumed to be sufficiently weak to neglect the effect of counter-rotating terms. The interactions in the Lamb-Dicke
expansion are given by
$$V_I=V_I^{(0)}+V_I^{(1)}( b + b^\dagger )+O(\eta^2)\,,$$ where the superscript gives the corresponding perturbative
order. The zero-order term reads~\cite{Zippilli05,Zippilli07}
\begin{equation}
\label{VI:0}
V_I^{(0)}= \hbar (\Omega+g  a^\dagger\cos \phi)\sigma + {\rm H.c.}\,,
\end{equation}
where $\Omega$ is the Rabi frequency, $g$ the vacuum Rabi frequency and $\phi$ the phase between a maximum of the cavity standing-wave and the center of the trap. Operators $\sigma=|g\rangle\langle e|$ and its adjoint $\sigma^{\dagger}$ flip the molecular excitation.

The term in first order in the Lamb-Dicke expansion include the coupling with the external degrees of freedom and read~\cite{Zippilli05,Zippilli07}
\begin{eqnarray}
\label{VI:1}
V^{(1)}_{I}  =  \hbar \eta (\mathrm{i}\Omega \cos\varTheta_L-g a^\dagger \cos\theta_C  \sin\phi )\sigma+ {\rm H.c.}\,,
\end{eqnarray}
where $\varTheta_L$ ($\varTheta_C$) is the angle between the axis of the motion and the laser (cavity) wave vectors.

The superoperators, describing dissipation in Eq. (\ref{Eq:master}), take the form
\begin{eqnarray}
&&{\cal L}_\kappa \rho = \frac{\kappa}{2} \left( 2a \rho a^\dagger-\{a^\dagger a, \rho\} \right)\,,\\
&&{\cal L}_\gamma \rho = \frac{\gamma}{2}
 \left( 2\sigma \tilde{\rho} \sigma^\dagger- \{\sigma^\dagger\sigma, \rho \} \right)\,,
\end{eqnarray}
with the molecular (cavity) linewidth  $\gamma$ ($\kappa$) and
$\tilde{\rho}$ accounting for the mechanical effects of the spontaneously emitted photon on
the motion~\cite{Zippilli05}.
Since spontaneous decay is very slow with respect to the other rates, in the simulations we
will take $\tilde{\rho}\simeq \rho$, hence discarding terms of the order $\eta^2\gamma$ which describe the diffusion due
to the recoil of the spontaneously emitted photons.

\subsection{Rate equations}

We characterize the efficiency of cavity-enhanced sideband cooling by means of a rate equation, which gives the dynamics of the occupation of the oscillator levels $|n\rangle$, $p_n={\rm Tr}\{|n\rangle\langle n|\rho\}$. This treatment is valid in the limit in which the dynamics of internal and cavity degrees of freedom is faster than the external motion, which requires $\eta\ll 1$. In this case, one can assume that the coupled system, given by the cavity mode and two-level transition,
reaches a steady state given by the density matrix $\varrho_{S}$, which is the solution of the equation $\mathcal
L_0\varrho_S=0$. Here, $\varrho_S$ is defined over the Hilbert space of cavity mode and internal degrees of freedom of
the molecule and 
\begin{equation}
\mathcal L_0\varrho=\frac{1}{{\rm i}\hbar}[H_M+H_C+V_I^{(0)},\varrho]+{\mathcal L}_{\gamma}\varrho+{\mathcal L}_{\kappa}\varrho
\end{equation}
is  the Liouville operator for internal and cavity degree of freedom. Under these assumptions the rate equation determining the dynamics of the trap states takes the form
\begin{equation}
\label{Eq:Rate}
\dot{p}_n=-(nA_-+(n+1)A_+)p_n+(n+1)A_-p_{n+1}+nA_+p_{n-1}\,,
\end{equation}
with~\cite{Zippilli05}
\begin{equation}
 \label{Eq:Apm}
 A_{\pm}=-2{\rm Re}\left[{\rm Tr}\left\{V_I^{(1)}\frac{1}{\mathcal L_0\mp i\nu}V_I^{(1)}\varrho_S\right\}\right]\,.
\end{equation}
Here, $A_+$ ($A_-$) is the heating (cooling) rate and $V_I^{(1)}$ is defined in Eq. (\ref{VI:1}). These are found by
assuming that the Rabi frequency $\Omega$ is the smallest parameter,  $\Omega \ll g,  \gamma, \kappa, \nu $, which
implies that the relevant internal and cavity states which are involved in the dynamics are the ground state
$|v=0,0_c\rangle$ and the excited states $|v=0,1_c\rangle, |v=1,0_c\rangle$. In the bad-cavity limit, state
$|v=0,1_c\rangle$ decays very fast, while $|v=1,0_c\rangle$ is characterized by a relatively smaller rate of decay, such
that $ \kappa \gg g \gg \gamma$ \cite{Zippilli05,Zippilli07}. An effective cooling rate $W$ can be defined for the case
$A_->A_+$ \cite{Stenholm86},
\begin{equation}
 W = A_- - A_+\,,
\end{equation}
which gives the rate at which phonons are exponentially damped towards the steady-state value of the trap. The steady-state expectation value of the phonon occupation number is given by \cite{Stenholm86}
\begin{equation}
 \langle n \rangle_{St} = \frac{A_+}{A_- - A_+}~.
\end{equation}

\subsection{Numerical evaluation of the master equation}
\label{sec:numerics}

We check the analytical predictions by means of numerical calculations, which consists in performing the time evolution
of the density matrix of cavity and molecule internal and external degrees of freedom according to Eq.
(\ref{Eq:master}). This is performed taking the Hilbert space of the cavity mode to be reduced to the photon states
$|0_c\rangle, |1_c\rangle$, which is a reasonable choice considering that in the bad-cavity limit the cavity mode is
either in the vacuum or has an occupation of one photon. Moreover, the molecular vibrational levels are approximated by
a two-level system, which is resonantly driven by radiation. Finally, the center-of-mass oscillator is approximated by
the first 5 levels. A convergence check for an increasing number of trap states has been carried out to ensure that the
error due to the truncation does not significantly affect the results. The resulting density matrix has dimensions
$20\times20$. The integration of Eq. (\ref{Eq:master}) is done with an fourth-order Runge-Kutta scheme to obtain the
time evolution of the coupled quantum system. The time evolution of the center-of-mass motion is found by taking the
partial trace of the total density matrix, as a function of time, over the internal and cavity degrees of freedom. In
order to determine the cooling rate, it is assumed that the rate equations, Eq. (\ref{Eq:Rate}), apply. Thus the time
evolution of the trap states is fitted against the populations $p_n$ to obtain $A_+$ and $A_-$, which allows us then to
extract $W$ and $\langle n\rangle_{St}$.

\section{Results}
\label{Sec:3}

In order to evaluate the efficiency of cavity sideband cooling of molecules, we first identify some good candidates.  We
note that a strong vibrational transition favors large
scattering into the cavity. Moreover, for the case of neutral molecules trapped by a far-detuned standing-wave laser a large polarizability is needed to generate a sufficiently deep potential well.
Several molecular candidates have been selected and calculated with the software package Gaussian \cite{Gaussian}. The equilibrium geometry of the molecules has been optimized with density functional theory and a high-quality atomic basis set (B3LYP/aug-cc-pVTZ \cite{Becke93, Dunning89, Dunning93}). Based on these geometries a frequency analysis and
polarizability calculation have been carried out to determine the spontaneous emission rates $\gamma$ of the excited
vibrational state and polarizability tensors as the decisive molecular properties. A small choice of possible candidates
which we have considered for implementing cavity sideband cooling is shown in
Tab.~\ref{tab:MolParam}.
\begin{table}
\caption{Parameters of the chosen molecules and their IR cooling transitions.
The point group of the molecule and the irreducible representation of the corresponding
vibrational normal mode is given. The transition frequencies, transition dipole moments($1\,\mathrm{au} =
2.54\,\mathrm{Debye}$) and
Einstein A coefficients $\gamma$ are approximated from the
{\it ab-initio} frequency analysis.}
\begin{tabular}{l|llrrr}
 	& PG &  Irred.      & $\tilde\nu_{12} [\mathrm{cm}^{-1}]$& $\mu_{12}[\mathrm{au}]$ & $\gamma[\mathrm{s}^{-1}]$
\\ \hline
CHBr$_3$   & $C_{3v}$	    & $E$      & 635  & 0.10 & 5.6 \\
HCCCF$_3$  & $C_{3v}$          & $A_1$    & 1234 & 0.15 & 80  \\
TMA        & $C_{3v}$          & $A_1$    & 2910 & 0.067& 226 \\
COS        & $C_{\infty v}$    & $\Sigma_g$ & 2108 & 0.15 & 424 \\
CFI$_3$    & $C_{3v}$          & $A_1$    & 1038 & 0.087& 17.1 \\
CSCl$_2$   & $C_{2v}$          & $A_1$    & 1131 & 0.13 & 5.12 \\
MgH$^+$    & $C_{\infty v}$    & $\Sigma_g$ & 1609 & 0.036  &11\\
\end{tabular}
\label{tab:MolParam}
\end{table}

The molecule COS is selected from the list as a neutral candidate for its remarkable strong IR transition in the
asymmetric stretch vibration.
Its large polarizability makes it also a good choice for optical trapping. For the two-level system $v=0$ and $v=1$ of the asymmetric stretch mode with the parameters given
in Tab.~\ref{tab:MolParam} are chosen.
The test candidate COS is assumed to reside in a standing-wave optical potential
which has a trap depth of $\approx 900\,\mu K$ and a trap frequency $\nu=2\pi\times350\,\mathrm{kHz}$ in the harmonic approximation. Such a potential could for example be formed by a laser
with a wavelength of 532\,nm and an effective power of 500\,W achieved by a build-up cavity.
The resulting Lamb-Dicke parameter is then $\eta = 0.02$. Molecular samples below 1\,mK may be achieved in the near future \cite{Zeppenfeld2009}, making it plausible to assume the Lamb-Dicke regime in our system.

The cavity axis and the pump laser axis is chosen to be in a 45\,$\mathring{}$ arrangement with the trap axis ($\Theta_C=\Theta_L=\pi/4$). The amplitude of the electric field of the cavity mode is $\varepsilon_c = 150\,\mathrm{Vm}^{-1}$ and the corresponding cavity line width is $\kappa = 2\pi\times5\,\mathrm{MHz}$. The single-particle cooperativity $C_1 = g^2/\kappa \gamma = 61$ warrants the strong-coupling regime which is needed to provide a sufficiently strong enhancement of the scattering rate. However, it is important to notice that the cavity coupling $g$ is smaller than the trap frequency ($g=0.41\,\nu$). This is a major difference to previous theoretical works \cite{Cirac95,Zippilli07} which assumed significantly larger values of $g$ with respect to $\nu$, and which determines in our case the disappearance of most interference phenomena reported in \cite{Cirac95,Zippilli07}.

In order to explore the parameter regime identified for the case of the COS molecule, the cooling and heating rates are
calculated by the two different methods described in Sec. \ref{sec:numerics}, namely, the analytical predictions from
perturbation theory and the numerical results found by integrating master equation Eq. (\ref{Eq:master}).
\begin{figure*}
 \includegraphics[width=0.98\textwidth]{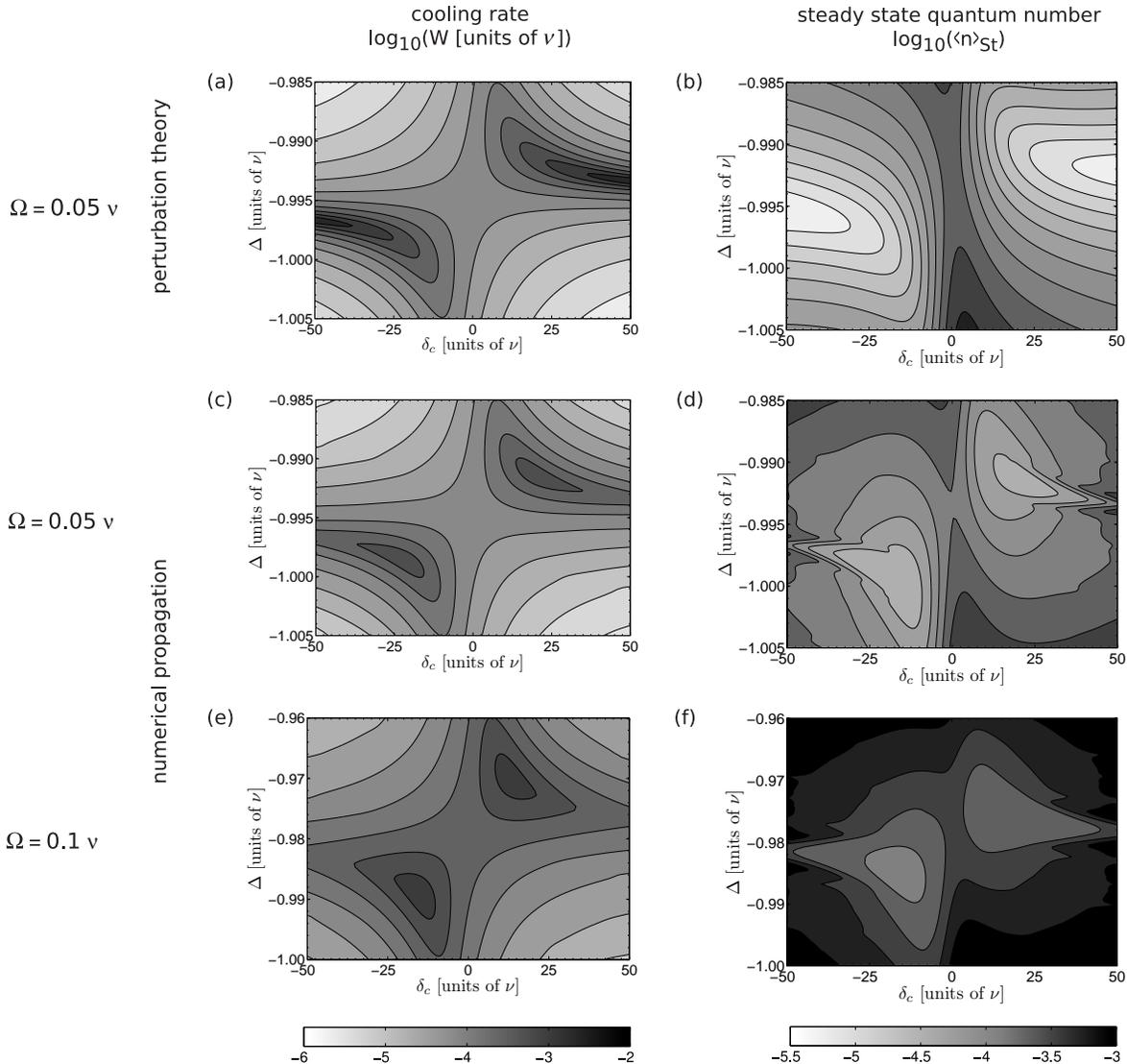}
 \caption{Contour plots of the cooling rate (left column) and corresponding steady-state occupation (right column) of
the motion of a trapped COS molecule driven at the asymmetric stretch vibration (2108\,cm$^{-1}$) as a function of  the
laser-molecule detuning $\Delta$ and the laser-cavity detuning $\delta_c$. The parameters are: $g=0.41\,\nu$, $\eta =
0.02$, $\kappa=14\,\nu$, $\gamma=1.9\cdot10^{-4}\,\nu$ and $\Theta_L = \Theta_C = \phi = \pi/4$. Subplots a) and b)
correspond to the results found by applying perturbation theory Eq. (\ref{Eq:Apm}) and taking $\Omega = 0.05\,\nu$.
Subplots (c) and (d) are found for the same parameters as in (a) and (b), but are calculated from the density matrix
found by numerical integration of Eq. (\ref{Eq:master}).
Subplots (e) and (f) are the same as (c) and (d) but with pump strength $\Omega = 0.1\,\nu$.}
\label{fig:COSW}
\end{figure*}

Figure \ref{fig:COSW} displays the contour plots of cooling rate and steady state phonon number as a function of
laser-molecule detuning $\Delta$ and the laser-cavity detuning $\delta_c$.
The top row in Fig. \ref{fig:COSW} shows to the predictions of the analytical model, while the second row reports the
corresponding predictions of the numerical results for the same set of parameters. Their comparison shows discrepancies
between analytics and numerics. This is understood: In fact, the perturbative treatment is based on the assumption that
the Rabi frequency  $\Omega$ is the smallest parameter, and that the excited state of the two-level transition is
essentially empty \cite{Zippilli07}, while for the parameters here considered with $\gamma$ being the
smallest rate this is not fulfilled.  We have also
checked that in the limit where the perturbative treatment is expected to be valid, the
predictions of analytics and numerics are almost identical. In the following we refer then to the numerical results. The
analytical results, nevertheless, serve for a double purpose. In first place, they allow one to identify the parameter
interval where cooling is optimal (allowing to quickly find the region where the numerical effort should be
concentrated on).
Moreover, since the dynamics described by the analytical solution is well understood, see e.g.
\cite{Zippilli05,Zippilli07}, by comparison  with the numerical results they permit one to identify the physical origin
of the discrepancies.

We first notice that efficient cooling is found for the values of the detuning at which the red sideband of the narrow
dressed state of cavity and molecule is driven. At large values of the detuning  $\vert \delta_c \vert \to \infty$, when
the coupling with the cavity mode becomes irrelevant, this dressed state tends to the molecular vibrational state $v=1$
and optimal cooling is thus found when $\Delta=- \nu$, which is the sideband cooling condition in free space
\cite{Stenholm86,Eschner03}. Correspondingly, the temperature is minimal. We also find that the numerical results
predict that the cooling rate exhibits a maximum (and the finite occupation a minimum) about a finite value of $\vert
\delta_c \vert$. Comparison with the analytical results shows that this behavior must be due to saturation effects,
which are not accounted for in the perturbative treatment. 
Indeed, for the chosen parameters the laser saturates the red sideband transition (being $\eta\Omega>\gamma$).
As a consequence, the mean vibrational occupation at steady state, extracted from the numerical method and displayed in
Fig. \ref{fig:COSW}(d), shows oscillations as the parameters are varied across the
sideband resonance.
The cooling rate which we extract at the maximum is $W=2\cdot10^{3}\,\mathrm{s^{-1}}$.

Figure \ref{fig:COSW}(e) and Fig.~\ref{fig:COSW}(f)  display the numerical predictions for cooling rate and steady
state occupation when the parameters are the same as in Figs. \ref{fig:COSW}(c) and Fig. \ref{fig:COSW}(d), but the
strength of the pump is doubled, $\Omega = 0.1\,\nu$. Larger cooling rates are achieved in this case, and the minimal
temperature is also increased. The observable power broadening leads also to a larger parameter space
where efficient cooling is observed.

\begin{figure}
 \includegraphics[width=0.45\textwidth]{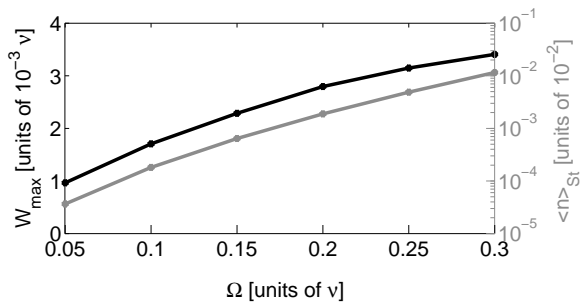}
  \caption{Extremal values of the cooling rate (black line) and of the steady state occupation (grey line) versus the
laser Rabi frequency (in units of $\nu$). The other parameters are the same as in Fig. \ref{fig:COSW}.}
 \label{fig:Max}
\end{figure}

Figure\,\ref{fig:Max} shows the maximum cooling rate as a function of $\Omega$, beyond the validity of the perturbative regime, which predicts that the cooling rate scales quadratically with $\Omega$ \cite{Zippilli07}. We observe that with increasing pump strengths the cooling rate becomes larger, and increases almost linearly with $\Omega$. Correspondingly,  the steady-state phonon number increases exponentially with $\Omega$. Thus the choice of
the pump strength is a trade-off between final temperature and fast cooling process.

\section{Conclusion}
\label{Sec:4}

In this article we have given numerical evidence that cavity-enhanced sideband cooling of trapped molecules is feasible,
by performing the calculations for a particular molecule, COS. The chosen example of COS has favorable properties such
as a
large polarizability and a strong vibrational transition. We have identified experimental parameters for trapping and
cooling the COS molecule, which may be realized in the near future. The use of infrared transitions well-resolved
sidebands may easily be achieved in different types of
harmonic traps. As long as the strong-coupling regime can be implemented
with a chosen vibrationally transition, concurrent decay channels are
outrun by the enhancement. Moreover the scheme is well suited for
ion traps due to their long storage times and their low anharmonicities. Under these considerations ground-state cooling
on a sub-millisecond time scale seems possible.
More efficient cooling could be achieved by increasing the the cavity
coupling $g$, which could be  achieved by exploiting collective enhancement in a molecular gas as proposed in Ref.
\cite{Lev08}.
Further improvement of the cooling process is conceivable by adaptive control of the Rabi frequency in time.

\begin{acknowledgements}
The authors acknowledge  support by the Spanish Ministry MICINN (QNLP FIS2007FIS2007-66944 and ESF-EUROQUAM CMMC
"Cavity-Mediated Molecular Cooling"), the German Research Foundation (MO1845-1/1), the European Commission (STREP
PICC, COST action IOTA) and the Deutsche Forschungsgemeinschaft through the excellence cluster Munich Centre for
Advanced Photonics
\end{acknowledgements}

\end{document}